\documentclass[twocolumn,prb,showpacs,superscriptaddress,preprintnumbers,amsmath,amssymb]{revtex4}

\usepackage[dvips]{graphicx}
\usepackage{hyperref,verbatim}
\usepackage{dcolumn,color}
\usepackage{bm}
\begin{document}

\title{Magnetic susceptibility of Cerium: an LDA+DMFT study}
\author{S.V. Streltsov}
\affiliation{Institute of Metal Physics, S.Kovalevskoy St. 18, 620041 Ekaterinburg GSP-170, Russia}
\affiliation{Ural Federal University, Mira St. 19, 620002
Ekaterinburg, Russia}
\email{streltsov@imp.uran.ru}

\author{E. Gull}
\affiliation{Department of Physics, Columbia University, New York, NY 10027}

\author{A.O. Shorikov}
\affiliation{Institute of Metal Physics, S.Kovalevskoy St. 18, 620041 Ekaterinburg GSP-170, Russia}
\affiliation{Ural Federal University, Mira St. 19, 620002 Ekaterinburg, Russia}

\author{M. Troyer}
\affiliation{Theoretical Physics, ETH Zurich, 8093 Zurich, Switzerland}

\author{V.I. Anisimov}
\affiliation{Institute of Metal Physics, S.Kovalevskoy St. 18, 620041 Ekaterinburg GSP-170, Russia}
\affiliation{Ural Federal University, Mira St. 19, 620002
Ekaterinburg, Russia}

\author{P. Werner}
\affiliation{Department of Physics, University of Fribourg, 1700 Fribourg, Switzerland}
\affiliation{Theoretical Physics, ETH Zurich, 8093 Zurich, Switzerland}

\begin{abstract}
The magnetic properties of Ce in the $\alpha$ and $\gamma$ phase are calculated within the LDA+DMFT
approach. The magnetic susceptibility in these two phases shows a similar behavior
over a wide temperature range: a Curie-Weiss law at high temperatures, indicating the presence of local moments, 
followed by a maximum in a crossover regime, and a saturation characteristic of a state with screened local moments 
at low temperature. The difference in experimentally observable magnetic properties 
is caused by the shift of the susceptibility to higher temperatures in the $\alpha$-phase 
compared to the $\gamma$-phase. 
\end{abstract}


\maketitle

\section{Introduction}
The isostructural $\alpha-\gamma$ transition in Ce is one of the
classical problems in modern solid states physics. In the
low temperature $\alpha$-phase (up to T$\sim 100$~K at normal conditions, or up to 
T$\sim 300$~K for a pressure $P$ of $1$~GPa) Ce behaves like a Pauli paramagnet,
while in the high temperature $\gamma$-phase the susceptibility
approximately follows a Curie-Weiss law.~\cite{Koskenmaki78} The transition
is accompanied by a drastic volume collapse (9-15\%)~\cite{Koskenmaki78} and
dramatic changes of the electronic spectra.~\cite{Liu92}

A number of theoretical models were proposed to
describe the $\alpha-\gamma$ transition. One of the first was
a promotional  model, where localized 4$f$ electrons were suggested to
transfer to the $spd-$(valence) band state, losing  their local moments.~\cite{Schuch1950,Lawson1949}
This was in contradiction to later experimental results that showed that the number of 4$f$ electrons is 
almost unchanged during the transition.~\cite{Gustafson69} 
As a result, a Mott-like picture was proposed, where the valence
of the Ce ions does not change, but the transition, which affects 
the degree of 4$f$ electron localization, occurs
as a result of the change of the ratio of on-site $f-f$ Coulomb interaction ($U$)
to kinetic energy.~\cite{Johansson74} 

Further neutron experiments~\cite{Murani05} confirmed that the Ce-4$f$ electrons
remain localized and indicated that the Kondo volume collapse
model~\cite{Allen82,Lavagna82} could be more plausible. According to this model
the 4$f$ electrons remain localized in the low temperature $\alpha-$phase, but
form Kondo singlets with conduction electrons in the $spd-$band. As a result
of this strong coupling the local spin moments of the 4$f$ electrons get screened, 
which leads to the volume collapse in the Kondo regime.

Thus a full account of the hybridization between localized 4$f$ and band $spd$
electrons is need for the correct description of Ce. Moreover, it was
recently shown that $f-f$ hopping is also important and should
be taken into consideration.~\cite{Streltsov10} The full
information about the non-interacting band structure of Ce can be obtained in the
frameworks of the density function theory (DFT), e.g. in the Local density approximation (LDA).
These density functional calculations can be extended to include local 
correlations within the LDA+DMFT scheme (combination of 
local density approximation and dynamical mean-field theory).\cite{Anisimov97}

``Ab-initio'' calculations in the LDA+DMFT approach 
were successfully applied to the modeling of electronic 
and structural properties of Ce. They have illustrated the key role of the entropic 
contribution to the free energy~\cite{Amadon06,Medici05} and the importance of 
the formation of a quasi-particle peak~\cite{Held01,McMahan03} for the 
description of the $\alpha-\gamma$ transition in Ce. The applicability of the 
Kondo model was mainly discussed via an analysis of the spectral properties
of the two phases: the temperature dependence of the quasi-particle 
resonance~\cite{Zofl01,McMahan03} or features of the Ce-$spd$ 
bands~\cite{Haule05}. Meanwhile, the description of a key  physical observable,
namely the temperature dependence of magnetic susceptibility
(for which different models were originally proposed) has not yet been attempted.

In the present report we use the LDA+DMFT method to calculate the magnetic 
properties of Ce in the $\alpha$ and $\gamma$ phases. 
The results of the study show that the temperature dependence of the
magnetic susceptibility is very similar in Ce$-\alpha$ and Ce$-\gamma$ and 
that both phases should thus be described by the same model.
The difference in the observed
magnetic properties is attributed to the decrease of 
the hybridization between localized $f$ and conductive $spd$
electrons, which leads to a shift of the magnetic
susceptibility curve to lower temperatures in Ce$-\gamma$.

\begin{figure}[t!]
 \centering
 \includegraphics[clip=false,width=0.4\textwidth,angle=270]{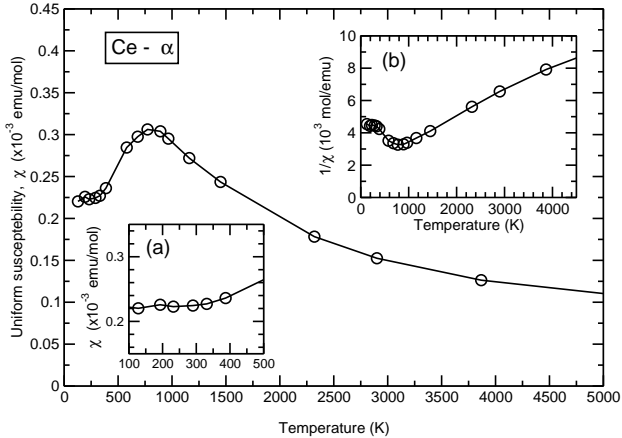}
\caption{\label{Chi-alpha} Uniform magnetic susceptibility $\chi (T)$ for Ce$-\alpha$. 
Inset (a) - enlarged view of $\chi (T)$ in the low temperature region. Inset 
(b) - inverse magnetic uniform susceptibility $\chi^{-1} (T)$.}
\end{figure}

\section{Calculation details}

We performed LDA calculations using the Linearized
muffin-tin orbitals (LMTO) method.~\cite{Andersen84}
An almost orthogonalized version of the LMTO in the
$\Gamma-$representation with Ce - 6s, 6p, 5d, and 4f states 
included to the basis set was used.
The Hamiltonian was generated on a mesh of 1728 $k-$points in the full Brillouin zone (BZ). 
This LDA hamiltonian was then transformed to a 
basis set with a diagonal form at the $\Gamma$-point. 
In this basis set the three lowest energy states at the $\Gamma$ point
correspond to the $t_{1u}$, the next three to the
$t_{2u}$ and the highest energy states to the $a_{2u}$ irreducible
representation.

The on-site Coulomb repulsion parameter ($U$) was estimated to
be 6.0 eV using a constrained supercell
calculation. This is in agreement with previous results.~\cite{Anisimov91G}
The intra-atomic Hund's rule coupling 
was set to $J_H$=0 eV.

For the solution of the DMFT equations we employed a diagrammatic 
(`continuous-time CT-HYB') quantum Monte Carlo algorithm which samples 
the partition function in powers of the impurity-bath hybridization.~\cite{Werner06,QMCReview} 
The Coulomb term was treated in the density-density form.
The double counting correction was set to $E_{dc}=U(n_\text{DMFT}-\frac12)$,\cite{Anisimov97} with 
$n_\text{DMFT}$ the total number of $4f$ electrons self-consistently
obtained within DMFT. The LDA+DMFT calculations was not fully self-consistent
in the sense that the LDA charge density was not recalculated
after the DMFT run. This can be done since the total number of
the $4f$-electrons doesn't change significantly.  
The Ce$-4f$ spectral functions were calculated using the maximum entropy method.\cite{Silver90}
We first computed the self-energy $\Sigma$ on the real frequency axis and used it to obtain the 
orbitally resolved and total spectral functions.

We calculated the uniform magnetic susceptibility $\chi$ as 
the ratio of the field-induced magnetization $d m(T)$ and the 
energy change $\delta E$ associated with the applied field ($h$)
\begin{equation}
\chi = \left.\frac {dm}{dh} \right|_{h \to 0} = 
\frac{n_{\uparrow} - n_{\downarrow}}{\delta E} \mu_B^2.
\end{equation}
Here $n_{\uparrow}$ and $n_{\downarrow}$ are the total occupation numbers
for spin up and down. The susceptibility $\chi$ was calculated for 
$\delta E=0.01$~eV, which is within the
interval where $m(T)$ is a linear function of $\delta E$.

\begin{figure}[t!]
 \centering
 \includegraphics[clip=false,width=0.4\textwidth,angle=270]{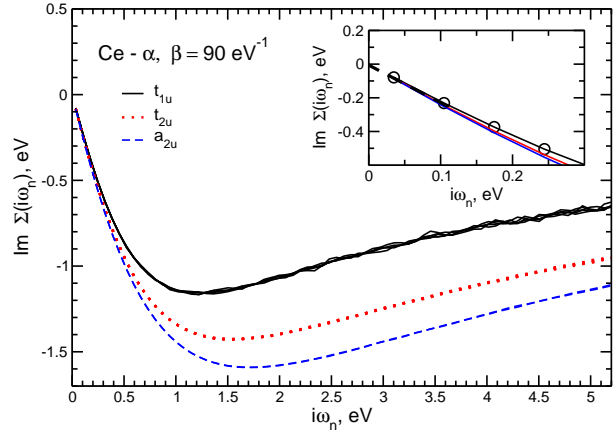}
\caption{\label{Sigma} (Color online) Imaginary part of the self-energy
for all 14 $4f-$orbitals at T=129~K ($\beta=90$~eV$^{-1}$) in Ce$-\alpha$. 
Due to the crystal-field splitting
they form three different sets of curves 
denoted as $t_{1u}$, $t_{2u}$ and $a_{2u}$.
The inset shows the low-energy behavior of $\text{Im} \Sigma (i\omega_n)$. 
For one of the curves exact positions
of Matsubara frequencies and linear extrapolation (by dashed line) to 
zero are shown.}
\end{figure}

\section{Results}

The uniform magnetic susceptibility $\chi(T)$ for the Ce-$\alpha$ phase, obtained using the LDA+DMFT calculations,
is shown in Fig.~\ref{Chi-alpha}. 
In the low temperature region the magnetic susceptibility is temperature
independent up to  $\sim$ 300~K. This is in qualitative agreement with 
experimental findings.\cite{Naka95} The absolute value of the magnetic
susceptibility is underestimated in LDA+DMFT by comparison with $\chi$ measured for P=1 GPa, where the 
presence of the structurally different 
Ce-$\beta$ phase is minimal: The experimental value of $\chi$ is $\sim 0.6-0.7 \times 10^{-3}$
emu/mol, while the theoretical estimate is $0.22\times 10^{-3}$ emu/mol. 
The underestimation
of the magnetic susceptibility could be due to the absence of long 
range correlations in our single-site DMFT calculations. Also, the spin-orbit coupling and associated orbital moment 
are not captured within our simple LDA description.

The plateau in the susceptibility of Ce at low-temperatures corresponds
to a coherent regime, where all 4$f$ electrons with local moments 
are screened. Indeed, one may notice from Fig.~\ref{Sigma} 
that in this temperature region the imaginary part of the self-energy for all $4f$ orbitals
is linear and approaches zero for low Matsubara frequencies. This is a signature that the
system at these temperatures is in a coherent Fermi-liquid 
regime.

\begin{figure}[t]
 \centering
 \includegraphics[clip=false,width=0.4\textwidth,angle=270]{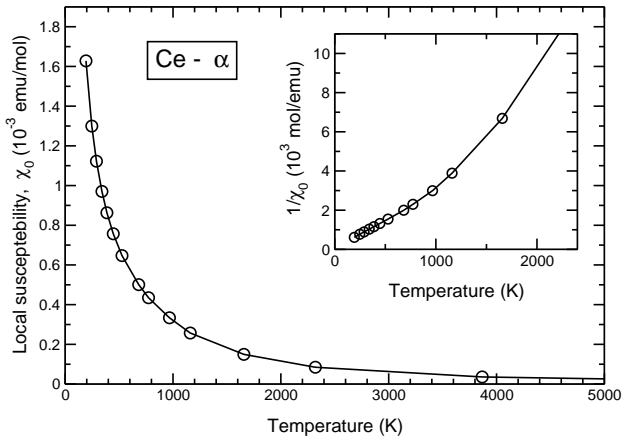}
\caption{\label{Local} Local magnetic susceptibility $\chi_0 (T)$ and inverse 
magnetic local susceptibility
$\chi_0^{-1} (T)$ for Ce-$\alpha$.}
\end{figure}

The importance of the screening effects by $spd$ electrons is most
clearly seen when one compares Fig.~\ref{Chi-alpha}, inset (a) and 
Fig.~\ref{Local}. The local magnetic susceptibility presented in 
Fig.~\ref{Local} was calculated as
\begin{eqnarray}
\chi_0 = \int_0^{\beta} \langle S_z(\tau)S_z(0) \rangle d\tau, 
\end{eqnarray}
where $\beta$ is the inverse temperature,
and $\langle S_z(\tau) S_z(0) \rangle$ is the imaginary time dependent 
spin-spin correlation function.
In spite of the strong non-linearity of the inverse 
uniform susceptibility, the local susceptibility follows a 
$\chi^{-1}_0(T) = T/C$ law up to T$\approx 1000$~K. 
This implies that the $4f$-electrons retain the 
local nature of the magnetic moment even at very low temperatures and there
is no delocalization process.
The violation of the Curie-Weiss law of the uniform susceptibility (Fig.~\ref{Chi-alpha}, inset (b)) and coherence 
of the system (which is seen in Fig.~\ref{Sigma}) is caused by the screening of local spins 
by conduction electrons.

With increasing temperature the screening effects weaken. This leads to an increase 
of the entropy according to Ref.~\onlinecite{Amadon06}.
Note that the growth of 
the susceptibility starts at $\sim$ 350~K, i.e. in the region where the
$\alpha-\gamma$ transition occurs.
\begin{figure}[t!]
 \centering
 \includegraphics[clip=false,width=0.4\textwidth,angle=270]{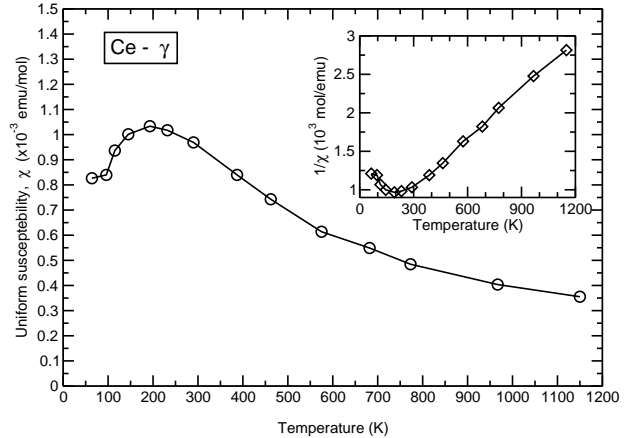}
\caption{\label{Chi-gamma} Uniform magnetic susceptibility $\chi (T)$ and inverse 
magnetic uniform susceptibility $\chi^{-1} (T)$ for Ce-$\gamma$, inset (a). 
}
\end{figure}
\begin{figure}[t!]
 \centering
 \includegraphics[clip=false,width=0.4\textwidth,angle=270]{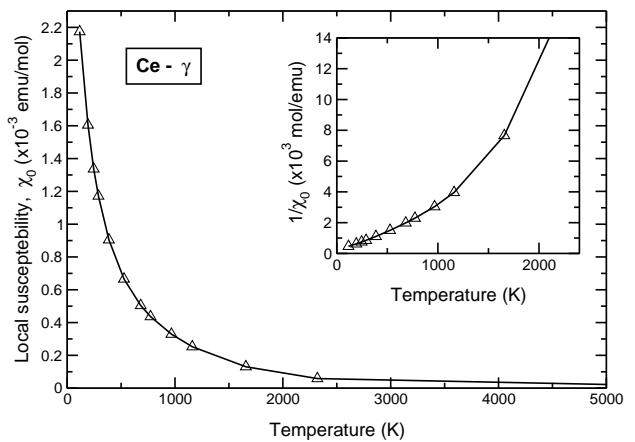}
\caption{\label{Local-g} Local magnetic susceptibility $\chi_0 (T)$ and inverse 
magnetic local susceptibility
$\chi_0^{-1} (T)$ for Ce-$\gamma$.}
\end{figure}

At temperatures above 350~K, the experimentally stable phase of Ce is the $\gamma$, not the $\alpha$ phase.
However, we may still simulate the magnetic 
properties of Ce$-\alpha$ even in a temperature region, where it does not exist by using
the LDA band structure of the corresponding phase. Making such
calculations one concludes that the
temperature range 350~K$<$T$<$1000~K is a crossover region, where the system
can neither be described as a coherent electron liquid nor as a lattice of localized $f$-states weakly hybridized to conduction electrons.
The maximum in the uniform magnetic susceptibility
at $820$~K appears as a result of a competition between Kondo 
and local spin regimes. Interestingly, in the 
Coqblin-Schrieffer model this maximum appears if the degeneracy of
the impurity level is more than 3.~\cite{Rajan83,Otsuki07} This is in agreement with the present
results, where the degeneracy of the lowest energy $t_{1u}$ states  
equals 6 (for both spins).  

A behavior consistent with the Curie-Weiss law, $1/\chi(T) \sim T$,  
for the uniform susceptibility is only seen above 
$1000$~K (Fig.~\ref{Chi-alpha}, inset (a)).

The uniform magnetic susceptibility in the $\gamma$ phase is qualitatively 
similar to the one just described for Ce$-\alpha$. A Curie-Weiss like behavior
at high temperature, then a crossover
region characterized by a maximum in the susceptibility, followed by a drastic drop and 
(at the lowest accessible temperatures) a saturation. 
The difference is mainly in the numbers. 
Already at T=$300$~K, the uniform magnetic susceptibility of 
Ce in the $\gamma$-phase is Curie-Weiss like. 
The broad maximum marking the crossover region is 
at $\sim 200$~K, much lower than in Ce$-\alpha$. 
However, since the $\gamma-$phase experimentally exists only above 300~K
the crossover regime and the constant $\chi(T)$ regime found 
below $\sim$90~K are not accessible in measurements.

The absolute value of the calculated 
$\chi$ in the region around 500~K, where it is experimentally measurable,
is underestimated (like in the $\alpha-$phase). We find $\sim0.7 \times 10^{-3}$ 
emu/mol against the experimental $\sim1.4 \times 10^{-3}$  emu/mol~\cite{Naka95}. 
As one may see in Fig.~\ref{Local-g} the local magnetic susceptibility, $\chi_0$, 
for Ce$-\gamma$ is similar to that for Ce$-\alpha$
and follows a $C/T$ law up to $T < 1000$~K. 

At higher temperatures $\chi_0^{-1}$ shows a  
deviation from the linear behavior with an upturn of the curve. 
A similar upturn can be observed for the inverse susceptibility 
of the one-band Hubbard model in the correlated metal regime 
(U=2.5, W=2).~\cite{StreltsovPriv12} 
This effect is due to the fact that
at higher temperatures the thermal fluctuations become so large that
the spins at different $\tau$ points become uncorrelated 
and ``do not feel" each other, i.e. that 
$\langle S_z(\tau)S_z(0) \rangle \sim C \to 0$
with $T \to \infty$. The result is a 
deviation of the local susceptibility $\chi_0^{-1} = T/C(T)$ from 
the Curie law.

\begin{figure}[t!]
 \centering
 \includegraphics[clip=false,width=0.4\textwidth,angle=270]{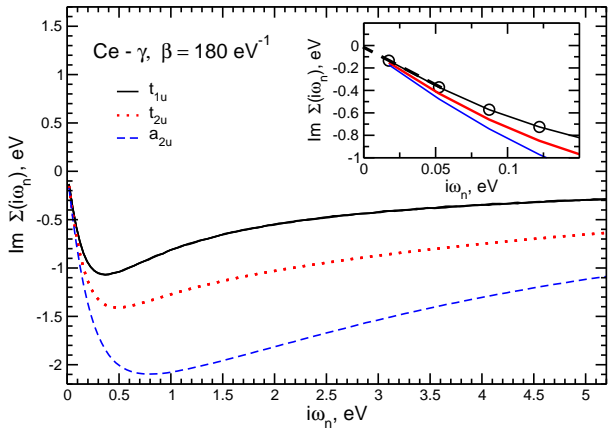}
\caption{\label{Sigma-gamma} (Color online) Imaginary part of the self-energy
for all 14 $4f-$orbitals at T=64~K in Ce$-\gamma$. 
Due to the crystal-field splitting
they form three different sets of curves 
denoted as $t_{1u}$, $t_{2u}$ and $a_{2u}$.
The inset shows the low-energy behavior of $\text{Im} \Sigma (i\omega_n)$. 
For one of the curves exact positions
of Matsubara frequencies and linear extrapolation (by dashed line) to 
zero are shown.
}
\end{figure}

\section{Discussion}

The shape of the susceptibility curves for the $\alpha$- and
$\gamma$-phases is very similar. Moreover, it strongly
resembles $\chi(T)$ obtained in models with a projected Hilbert space, such
as the Coqblin-Schrieffer impurity,\cite{Rajan83,Otsuki07} its lattice version,\cite{Otsuki09} 
and also $\chi(T)$ in the Periodic Anderson model.\cite{Leder79} The uniform static magnetic
susceptibility in all these models is characterized by a high-temperature
Curie-Weiss tail, a maximum in the intermediate regime and then a drastic
drop with a constant susceptibility at the lowest temperatures.

\begin{figure}[t!]
 \centering
 \includegraphics[clip=false,width=0.4\textwidth,angle=270]{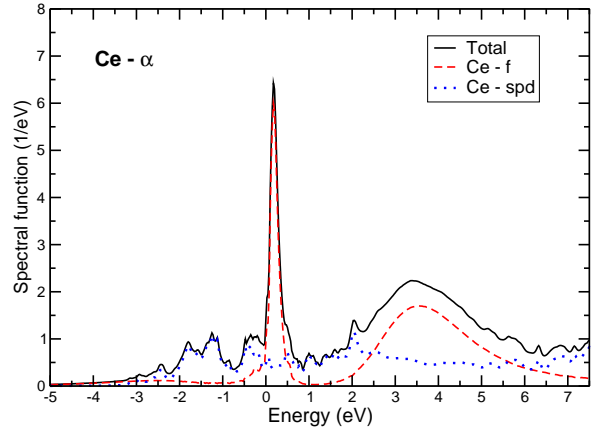}
 \includegraphics[clip=false,width=0.4\textwidth,angle=270]{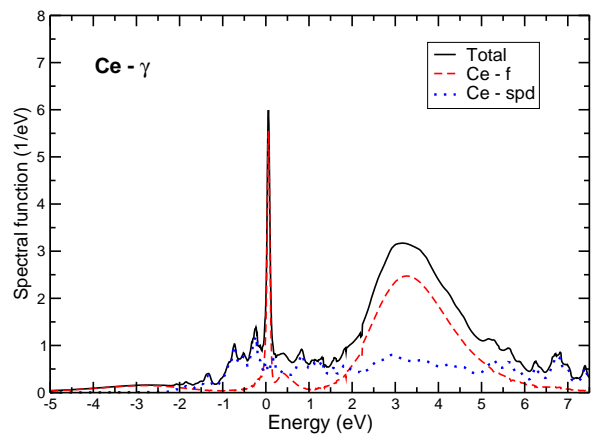}
\caption{\label{DOS} (Color online) Top panel: Spectral function for the Ce$-\alpha$
phase calculated for $\beta=90$ eV$^{-1}$.
Bottom panel: Spectral function for the Ce$-\gamma$
phase calculated for $\beta=180$ eV$^{-1}$
}
\end{figure} 

The high-temperature behavior of $\chi(T)$ is explained by the 
presence of local moments and described in a similar manner in all of the 
models, while the drastic decrease of the magnetic susceptibility 
at intermediate temperatures 
is caused by the screening of this moment. 
The lattice effects such as non-zero hopping between localized and
conductive states, centered at different sites are already taken into account 
in our LDA+DMFT calculation.
The results clearly show the formation of a coherent 
state with $\text{Im} \Sigma (i\omega_n) \to 0$ at low temperatures 
(see Fig.~\ref{Sigma},\ref{Sigma-gamma}). 
Meanwhile, an inspection of the spectral functions, plotted in 
Fig.~\ref{DOS}, shows that 
there is no gap or pseudo-gap in the vicinity of the Fermi level in the $f$-electron spectral 
function. Note that we present here results at considerably lower temperatures than in 
previous DMFT studies.

Comparing Fig.~\ref{Chi-alpha} and~\ref{Chi-gamma} one sees 
that the behavior of the uniform magnetic susceptibility in 
$\alpha-$ and $\gamma-$Ce is qualitatively the same. The
susceptibility in the $\gamma$ phase seems to be shifted
to the low-temperature region and renormalized (as compared with
Ce$-\alpha$). Thus, one may argue that those phases are physically 
similar in a wide temperature range.

The change of the lattice volume under the $\alpha - \gamma$ transition 
results in a modification of the hybridization function in the vicinity 
of the Fermi level and decrease of the $f-f$ hopping. 
The weakening of the hybridization in Ce$-\gamma$ 
(see Ref.~\onlinecite{Zofl01} for instance) leads to a decrease of the 
exchange parameter $J_K$ in the Kondo model according to the 
Schrieffer-Wolff transformation.~\cite{Schrieffer66} 
Numerical calculations show that the decrease of the Kondo
exchange results in a shift of the maximum of
the magnetic susceptibility to lower temperatures.~\cite{Otsuki07}
This is exactly what is seen in Ce under the $\alpha - \gamma$ transition.
A similar shift of the maximum of $\chi(T)$ due to a change of the hybridization 
was also observed in the study of the magnetic properties of the 
Periodic Andersen model (PAM).~\cite{Leder79}

It is also instructive to compare the present results 
with the situation in Pu, which is on the border of the transition
between the actinide elements with localized and delocalized electrons.~\cite{Moore09} 
Moreover the degree of localization changes in
the different phases of Pu.~\cite{Baclet07,Havela02} The DMFT calculations of Pu 
show that the local magnetic susceptibility dramatically changes
from Pauli-like to Curie-Weiss as the volume increases.~\cite{Marianetti08} 
This is completely different from the situation observed in Ce, where
the local susceptibility follows a $1/T$ law both for the $\alpha$ and $\gamma$
phases. This is further evidence for the absence of any localization/delocalization
transitions in the $4f$-shell of Ce, i.e. against the Mott transition scenario.

To summarize, in the present paper the magnetic susceptibility
for the $\alpha$ and $\gamma$ phases of Ce was investigated. 
It exhibits a qualitatively similar behavior in both phases. 
There is no Mott transition for any of the Ce phases.
On the local level they are both characterized by the presence of magnetic moments, 
screened by band $spd$ electrons at low temperatures. With increase of the temperature this
coherent state with constant susceptibility is gradually destroyed by thermal excitations, which results in the 
formation of a Curie-Weiss paramagnetic state in the high temperature region.
The difference in experimentally observable magnetic properties for the two
phases of Ce is related to a shift of 
the susceptibility in Ce$-\gamma$ to lower temperature region
(as compared to $\alpha$ phase).

\section{Acknowledgments}
We thank S. Skornyakov and A. Poteryaev for the help
with the calculations of the local susceptibilities.
This work was supported by grants RFBR 10-02-00046 and 10-02-96011,
the program of the President of the Russian Federation MK-309.2009.2,
grant of the Ural branch of the Russian Academy of Science for young
scientists, SNF grant PP0022-118866 and NSF-DMR-1006282. 
The calculations were based on the ALPS \cite{ALPS} 
DMFT \cite{ALPSDMFT} package.

\bibliography{refs.bib}
\end{document}